# Spectrally Extended Line Field Optical Coherence Tomography Angiography

Si Chen, Kan Lin, and Linbo Liu*

*School of Electrical & Electronic Engineering, Nanyang Technological University, Singapore*

*Correspondence: L. L. (liulinbo@ntu.edu.sg)

**Abstract** Optical coherence tomography angiography (OCTA) has been established as a powerful tool for investigating vascular diseases and is expected to become a standard of care technology. However, its widespread clinical usage is hindered by technical gaps such as limited field of view (FOV), lack of quantitative flow information, and suboptimal motion correction. Here we report a new imaging platform, termed spectrally extended line field (SELF) OCTA that provides advanced solutions to the above-mentioned challenges. SELF-OCTA breaks the speed limitations and achieves two-fold gain in FOV without sacrificing signal strength through parallel image acquisition. Towards quantitative angiography, the 'frequency flow' imaging mechanism overcomes the imaging speed bottleneck by obviating the requirement for superfluous B-scans. In addition, the 'frequency flow' imaging mechanism facilitates OCTA-data based motion tracking with overlap between adjacent line fields. Since it can be implemented in any existing OCT device without significant hardware modification or affecting existing functions, we expect that SELF-OCTA will make non-invasive, wide field, quantitative, and low-cost angiographic imaging available to larger patient populations.

Optical coherence tomography angiography (OCTA) is a functional extension of optical coherence tomography (OCT), capable of noninvasively highlighting blood flow signals at the capillary level in three dimensions (3D). Since OCTA can obtain high quality, high-contrast angiograms quickly without the need for dye injection, this capability opens a wealth of possibilities for screening, surveillance and diagnosis of diseases, and development and evaluation of new treatments [1-6]. Over the past decade, its use has been rapidly expanding in clinical practice and research, particularly with the commercialization of ophthalmic OCTA techniques in the past several years. However, the widespread usage of OCTA is still dependent on overcoming challenges such as limited field of view (FOV), motion artifacts, and lack of flow velocity quantification.

OCTA relies on moving red blood cells to contrast microvasculature against static tissue. This contrast signal induced by blood flow is extracted by detecting variations of OCT signals between consecutive scans taken at the same location within a Y-scan cycle. A typical OCT device adopts the point-scanning mechanism, in which the beam scanners steer the beam so that the light spot dwells at one lateral position at a time. Because of the sequential nature of this mechanism, achievable lateral FOV is solely dependent on A-line rate. This dependency is especially acute in OCTA, since 2-5 repeated B-scans are required for motion contrast. As a result, FOV of most ophthalmic OCTA systems is much smaller than the standard-of-care technologies [1, 2], even though capillary fine details are always sacrificed due to under-sampling [1, 7]. Swept-source technologies are promising in achieving ultra-wide field OCTA with dramatically improved A-line rate, whereas, increases in imaging speed decrease the OCT signal and speed is ultimately limited by the signal to noise requirements given the constraint of allowable light exposure [2]. Therefore, the trade-off between FOV and signal quality is inherent with current OCTA, as long as both are solely dependent on the A-line rate.

Wide-field OCTA is not possible without effective tracking and correction of motion artifacts. For ophthalmic applications, eye tracking relies on additional imaging hardware such as infrared fundus camera or scanning laser ophthalmoscopy [8-10], which can extend the available imaging time beyond the few seconds when patients can fixate without saccades or blinking [2]. However, fundus images are not adequately sensitive to small motions due to their relatively low resolution. Since there is always a latency, there are errors that are not correctable by hardware-based eye tracking [2]. In addition, the increased system complexity and cost are also practical concerns. Self-navigated motion correction method represents a new trend to tackle these issues, which suppresses eye motion and blinking artifacts on wide-field OCTA without requiring any hardware modification [11]. However, none of the existing software-based techniques is able to track lateral motion, leaving significant artifacts uncorrected. Improvement in motion artifact management requires a new technology that is capable of tracking motion with high precision and minimal latency, and ideally without additional hardware.

Quantification of flow velocity is of great interest with regards to disease diagnosis and management [1, 2, 12-15]. It has been well understood that OCTA flow signal is primarily affected by inter-scan time, which is the time interval between repeated B-scans at the same position. If the inter-scan time is long, the OCTA signal saturates easily for higher range of velocities, so that relation between flow speed and signal is nonlinear and complicated [14, 16-18]. In contrast, a short inter-scan time can better distinguish higher range of velocities; however, sensitivity to slow flow is reduced, as the red blood cells do not have sufficient time to move far enough to produce a detectable signal variance [2]. Choi [12] and Wei [19] have managed to extend the dynamic range of detectable flow velocity by generating OCTA signals of multiple inter-scan time. However, these point-scanning based approaches requires more B-scan repeats than the standard OCTA, further limiting the FOV. A new technique capable of acquiring OCTA signals of multiple inter-scan time independent of the number of B-scan repeats is needed to have this important metric in routine applications.

The core of OCTA techniques is algorithms that computes OCT signal variations, which include, but not limited to, optical microangiography [20], speckle variance [21], phase variance [22], split-spectrum amplitude-decorrelation angiography (SSADA) [22] and correlation mapping [23]. Interestingly, split-spectrum and subsequent frequency compounding increase the image quality for the algorithms most commonly implemented for ophthalmic applications [24, 25]. The reason is that because speckle pattern is wavelength dependent, by splitting the full OCT spectrum into a number of narrower bands, frequency compounding reduces spectral-dependent speckle noise in flow-generated speckle signals [26]. The representative technique, SSADA, has been extensively validated in clinical settings [12, 14, 15, 27]. Splitting spectrum creates a new dimension for parallel image acquisition. Spectrally encoded OCT have been proposed to achieve parallel structural image acquisition [28]. However, towards high-quality blood flow imaging, how different frequency components can be compounded at the same image position is not known.

We report a novel imaging platform, termed spectrally extended line field OCTA (SELF-OCTA), which achieves frequency-temporal compounding through a 'frequency flow' imaging mechanism. SELF-OCTA provides advanced solutions to the above-mentioned challenges towards significantly improved clinical utility of OCTA.

**Method**

We have developed a spectral-domain OCT (SD-OCT) device working at the center wavelength of ~1310 nm, which allows convenient switching between the SELF scheme and the standard point-scanning scheme (Fig. 1A). The construction of the imaging system is mostly the same as a typical OCT device except a set of prisms in the sample arm. Detailed information on system construction is provided in ***Supplementary Information***. With a collimating lens of 10 mm focal length and objective lens of 50 mm focal length, the theoretical lateral spot size at 1310 nm is ~27 µm (full-width at half maximum, FWHM). SELF-OCTA achieves 2D (depth and slow axis) priority scan by spectrally extending the polychromatic beam to a line field along Y-axis in the focal plane of the objective lens (Fig. 1B).

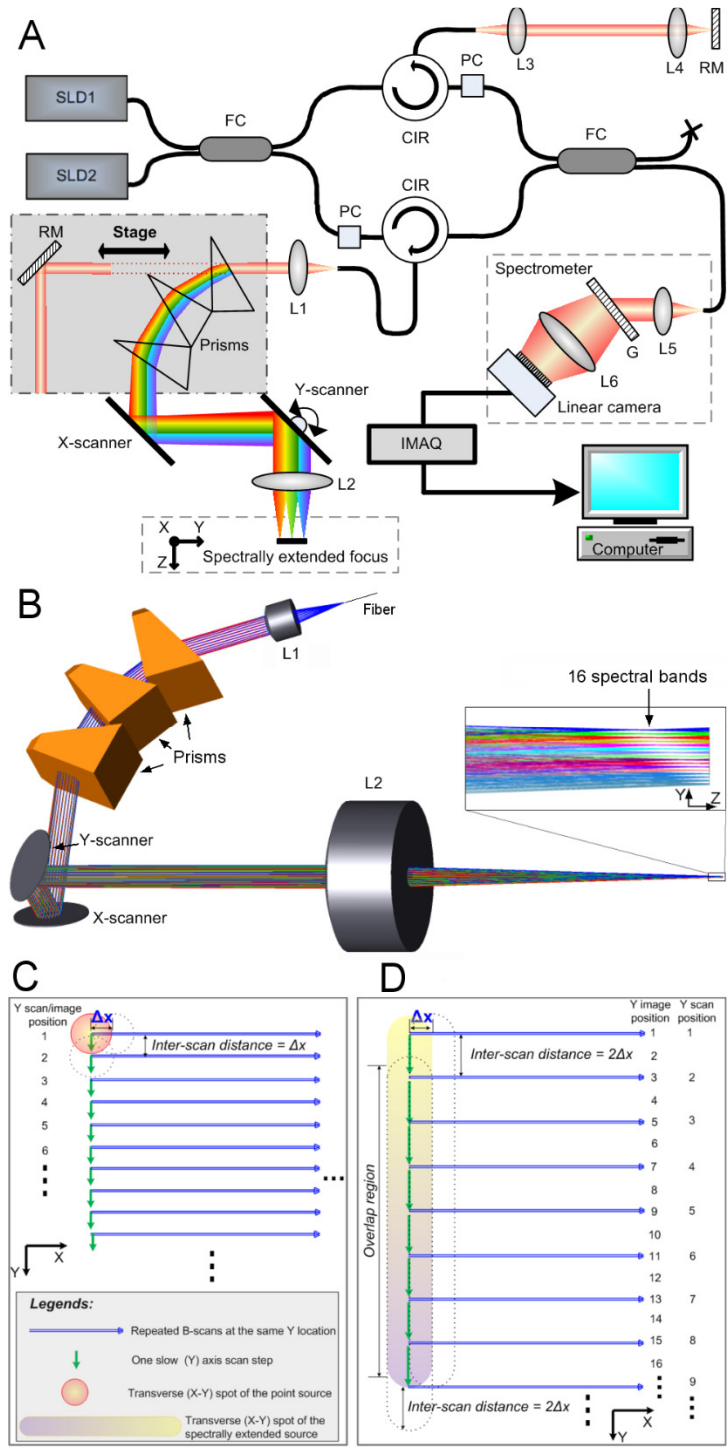

**Figure 1.** (**A**) System schematic of SELF-OCTA. SLD1&SLD2: superluminescent diode sources; FC: fiber coupler; PC: polarization controller; CIR: circulator; L1-5: achromatic lenses; L6: camera lens; RM: reference mirror; G: transmission grating; IMAQ: image acquisition card. (**B**) Three-dimensional layout of the sample arm optics and spectrally extended line field (inset). (**C**) Beam scanning protocol of the point-scanning scheme. Δx is the step size along the X-axis and the inter-scan distance is the spacing between two concecutive Y-scan positions. Dashed circles represent spots of adjacent X/Y-scan positions, respectively. (**D**) Beam scanning protocol of SELF-OCTA. Dashed contours represent line fields of adjacent X/Y-scan positions, respectively. Note that the Y-scan position is the Y position illuminated by the first spectral band during each Y-scan cycle, and is different from the Y image position.

In this study, we imaged skin vasculatures at the palm side of the proximal interphalangeal (PIP) joint of the middle finger with healthy human subjects. This study was approved by the Institutional Review Board (IRB) of Nanyang Technological University (IRB-2016-10-015 and IRB-2019-05-050). For the experiments conducted with the point-scanning scheme and total image acquisition time of 4.096 s (Fig. 3 and **Supplementary Fig. 7**), the optical power incident on the sample was 4.74 mW. For the experiments conducted with the point-scanning scheme and total image acquisition time of 3.26 s, the optical power incident on the sample was 9.1 mW (Fig. 4). We used ~9.10 mW for all the experiments with the SELF scheme. The optical power incident on the skin is below American National Standards Institute (ANSI) exposure limit for skin safety.

Since the thermal damage threshold of an extended source is higher than the corresponding point source [29-31], higher maximum permissible exposure (MPE) is allowed in favour of signal strength [32]. The optical power limit in the SELF scheme, is calculated using the correction factor for exposure limit of extended sources for ocular safety [32, 33]. By applying the 'Most Restrictive Ratio' method [34, 35], the angular subtense in Y-axis generated by the prisms is ~3.176 mrad and the normalized partial power within the angular subtense is 0.799 (**Supplementary Fig. 1**). The corrected optical power incident on the skin in the SELF scheme is 9.25 mW. Detailed method to determine the corrected sample power is provided in **Supplementary Information**.

The X-axis scanning protocol is identical between both schemes: the scanning step size Δx is set to be 12.8 μm to satisfy the Nyquist sampling requirement, and the number of repeated B-scans at the same location $N$ equals to 2. The Y-axis scanning protocol, however, may be different. For the ease of comparison, we define Y-scan positions as Y positions illuminated by the first spectral band during each Y-scan cycle (Figs. 1C&D). In the point-scanning scheme, the inter-scan distance, which is the distance between two adjacent Y-scan positions, is equal to the X-axis scanning step size Δx, that is, Y image positions are the same as Y-scan positions (Fig. 1C). We followed SSADA algorithm for processing data acquired from the point-scanning system [22], in which we split the source spectrum into $M$ = 16 equally spaced bands in wavenumber domain using Hamming filters (**Supplementary Fig. 2A** and **Supplementary Information**), and all $M$ partial-spectrum decorrelation frames are averaged to obtain the OCTA signal at one Y image position. In SELF-OCTA, we also split the source spectrum into $M$ = 16 equally spaced bands using the same Hamming filters. The spacing between adjacent bands is Δy = Δx, so that each spectral band samples a distinct Y image position (Fig. 1D and Fig. 2A). The positioning error caused by the nonlinear frequency-space relation is negligible (**Supplementary Fig. 3**). By setting the inter-scan distance to be $L$·Δx ($L$ = 2, 4, 8, or 16), achievable FOV is multiplied by a factor of $L$. In general, $M$ spectral bands acquired at $j$-th Y-scan position illuminate $M$ consecutive Y image positions: ($j$-1)·$L$+1, ($j$-1)·$L$+2, …($j$-1)·$L$+$M$, respectively. In other words, $M/L$ light beams of distinct spectral bands dwell at the same lateral image position sequentially in time during Y-axis scan. This is analogous to the flow production process, where a number of Y image positions are addressed in parallel, and frequency components at each image position is 'assembled' during a number of consecutive Y-scan cycles. We term this slow axis imaging mechanism 'frequency flow'.

The split-spectrum data is processed to generate $M$ partial-spectrum decorrelation frames through Discrete Fourier transform (DFT) and amplitude decorrelation, which is similar to SSADA [22] (Fig. 2A). In general, $m$-th ($m$ = 1,2…$M$) partial-spectrum decorrelation frames acquired at $j$-th Y-scan position are assigned with an index number of $M$·($j$-1)+ $m$·$M/L$-quotient[($m$-1)/$L$]. By doing so, the final OCTA signal at $i$-th Y image position is the average of partial-spectrum decorrelation frames with index number from ($i$-1) ·$M/L$+1 to $i$·$M/L$. Figs. 1D&2B illustrate a scanning protocol with $L$ = 2, $M$ = 16 and $N$ = 2, where 16 partial-spectrum decorrelation frames generated from spectral interference data acquired in $j$-th Y-scan position contribute to OCTA signals of 16 consecutive Y image positions. For example, in Fig. 2B the final decorrelation frame at the Y image position of $i$ = 15 is obtained by averaging 8 partial-spectrum decorrelation frames: 15th frame at 1st Y scan position, 13th frame at 2nd Y scan position, 11th frame at 3rd Y scan position, 9th frame at 4th Y scan position, 7th frame at 5th Y scan position, 5th frame at 6th Y scan position, 3rd frame at 7th Y scan position and 1st frame at 8th Y scan position.

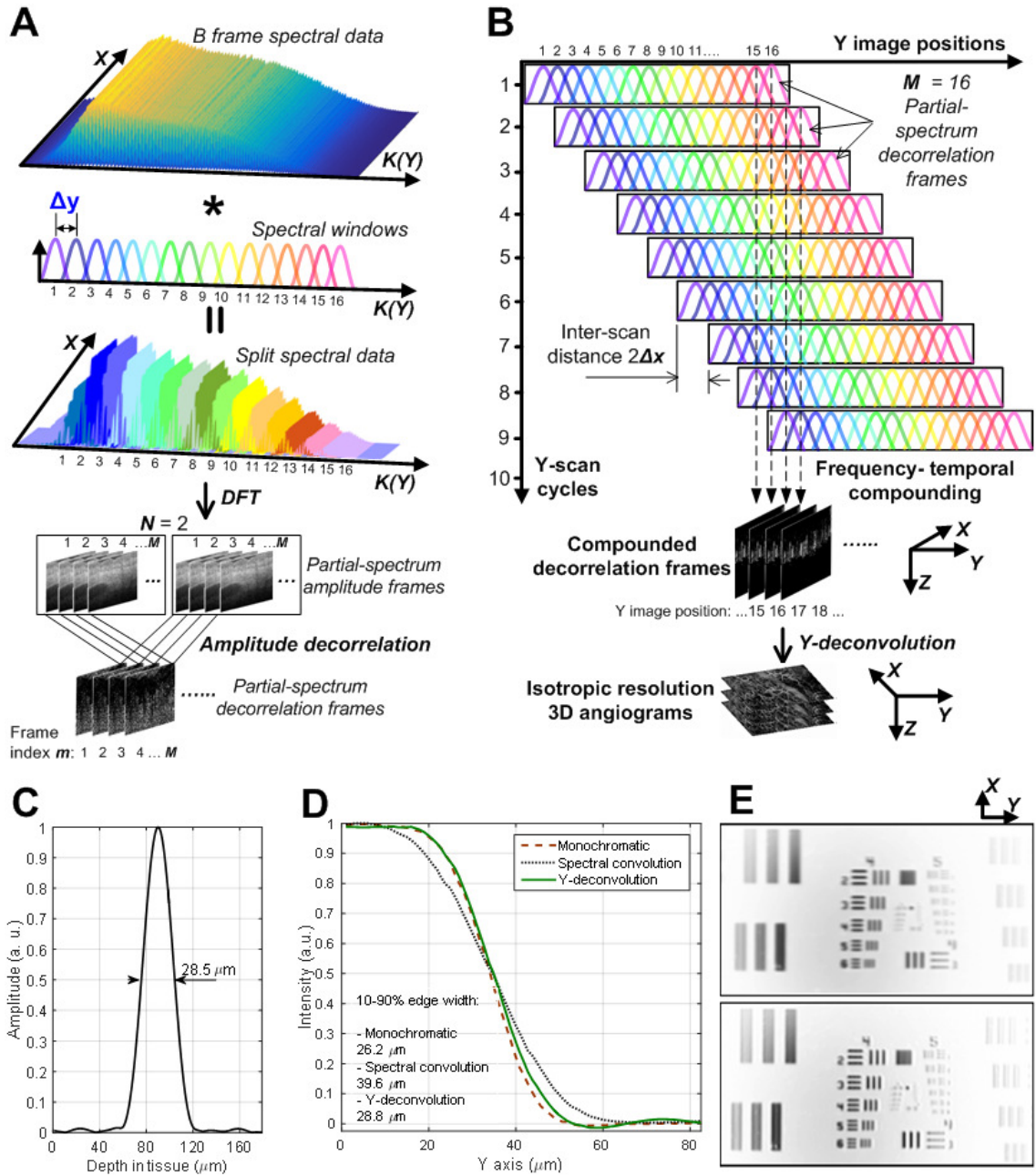

**Figure 2**. (**A**) Process of spectrum splitting and amplitude decorrelation. ***K*** is the wave-number and Δy is the spacing between Y image positions and equals to Δx. DFT: discrete Fourier transform. Spectral windows are hamming filters in this study. (**B**) "Frequency flow" imaging mechanism with ***M*** = 16 and ***L*** = 2. In the first Y-scan cycle, 16 spectral bands of the line field dwell at Y image positions of 1 to 16, respectively. In each Y-scan cycle, the image position numbers covered by the line field increase by ***L*** = 2. For a given Y image position, the final OCTA signal (compounded decorrelation frame) is the average of 8 (***M/L***) partial-spectrum decorrelation frames that are sequentially acquired from the Y image position. (**C**) OCT depth profile of a partial reflector as the sample. (**D**) 10-90% edge scan profiles of a monochromatic beam and the beam of the eighth spectral band (convolution in Y-axis). The green solid profile (Y-deconvolution) is obtained by deconvoluting the latter along Y-axis. (**E**) *En face* images of USAF 1951 resolution chart obtained using the beam of the eighth spectral band before (upper panel) and after Y-deconvolution (lower panel).

The Hamming window length is ~39 nm (***Supplementary Figs. 2A***), so that the axial resolution is measured to be 28.5 µm in tissue with refractive index of tissue of 1.38 (Fig. 2C). We model the multiplication between interferometric spectral data and a Hamming filter as a convolution in Y-axis, that is, the polychromatic lateral point-spread function along Y-axis is the convolution between the monochromatic point-spread function (PSF) and the Hamming window (**Supplementary Figs. 2B&C**). The lateral resolution is characterized by use of the 10-90% width of an edge scan profile as well as imaging a resolution chart (Figs. 2D&E). The lateral PSF in Y-axis is broadened to 1.51 times of the monochromatic PSF due to the convolution in Y-axis, which agrees well with the model (**Supplementary Fig. 2C**). We use a one-dimensional deconvolution algorithm to restore the lateral resolution along Y-axis in the *en face* projections (Fig. 2B). The 10-90% edge width restored by deconvolution is comparable to that of monochromatic light (Fig. 2D), so that isotropic spatial resolution is achieved in SELF-OCTA, which are also corroborated by the *en face* images of resolution chart (Fig. 2E). After deconvolution (lower panel, Fig. 2E), group 4 element 5 with line width of 19.69 µm can be unequivocally resolved in both X and Y directions, and Y-axis resolution is comparable to that of X-axis, which is the point-scanning direction. Details of Hamming filters, deconvolution, and OCT structural image processing are provided in **Supplementary Information**.

Images were acquired at an A-line rate of 50,000 Hz and 512 A-lines per B-frame for both schemes unless otherwise specified. For side-by-side comparison, we imaged the same skin region by careful alignment using cross-sectional OCT previews, except for **Supplementary Fig. 7**. We used home-made vertical and horizontal hand rests to minimize motion. We manually segmented the cross-sectional angiograms into three layers according to image depth with respect to the top of capillary loops (Figs. 3A-D): the first layer is ~180 µm thick in tissue, which includes most of capillary loops; the second layer is also ~180 µm thick in tissue, which mainly depicts subpapillary plexus; the third layer is ~360 µm thick in tissue, which corresponds to the skin layer with deep vascular plexus. Detailed information on image acquisition and image format are provided in **Supplementary Information**.

In addition, to demonstrate motion tracking and correction, we acquired OCTA images while deliberately moving hands along the lateral directions (Fig. 6). Before image acquisition, we immobilized a short segment of piano wire (F1-8265, Fiber Instrument Sales, Inc.) to the skin surface with the refractive index matching gel. The wire was aligned along Y-axis or oblique with respect to Y-axis so that it served as the reference for lateral motion (Fig. 6A). As mentioned above, a 3D OCTA dataset is acquired during each Y-scan cycle with the SELF-OCTA scheme, which allows us to track lateral motion using *en face* projects of these 3D angiograms. Since each *en face* projection is 512 × 16 pixels (X × Y) in size, with $L$ = 2 there is an overlap of 14 pixels along Y-axis between projects of adjacent Y-scan cycles (Fig. 6B). We measured lateral displacement between the overlap portions by 2D cross-correlation [36]. Note that we excluded the piano wire image from the input of the motion tracking. Details of motion tracking and correction algorithms are provided in **Supplementary Information**.

**Results**

We compared FOV between two schemes under the same conditions: 512 A-lines per B-frame, 400 Y-scan positions, and a total acquisition time of 4.096 s (Fig. 3). The FOV scanned with the point-scanning scheme is 6.55 mm x 5.12 mm (Figs. 3E1-4). SELF-OCTA provides twice as large FOV when we doubles the inter-scan distance as illustrated in Fig. 1D (Figs. 3F1-4&G1-4). With the same display contrast, the image quality of SELF-OCTA is comparable to that of the point-scanning scheme. First, through one-to-one comparison the vascular microstructures are almost identical between the point-scanning and SELF-OCTA images, including the capillary loops (Figs. E2, F2&G2). The SELF-OCTA *en face* images are slightly less crispy before deconvolution because of the convolution effect mentioned above (Figs. 3F1-4). Nevertheless, this insignificant issue is corrected after deconvolution (Figs. 3G1-4). Secondly, the penetration depth is also comparable as can be seen in the *en face* images of deep vascular plexus (Figs. E4, F4&G4), and cross-sectional angiograms (Figs. 3B-D and **Supplementary Fig. 4**). Thirdly, the mean decorrelation, measured from one-to-one matched vascular areas (**Supplementary Fig. 5**), are also comparable between the point-scanning OCTA (0.206 ± 0.006) and SELF-OCTA (0.204 ± 0.004) (Fig. 3H) (student's t-test, $p$ = 0.23). The speckle contrast of *en face* SELF-OCTA images (0.439 ± 0.059) is close to that of the point scanning scheme (0.353 ± 0.036), although the number of partial-spectrum decorrelation frames to be averaged at each Y image position is half of that of the point-scanning scheme. We attribute this relatively low speckle contrast to the fact that the partial-spectrum decorrelation frames at a Y image position are acquired at different time.

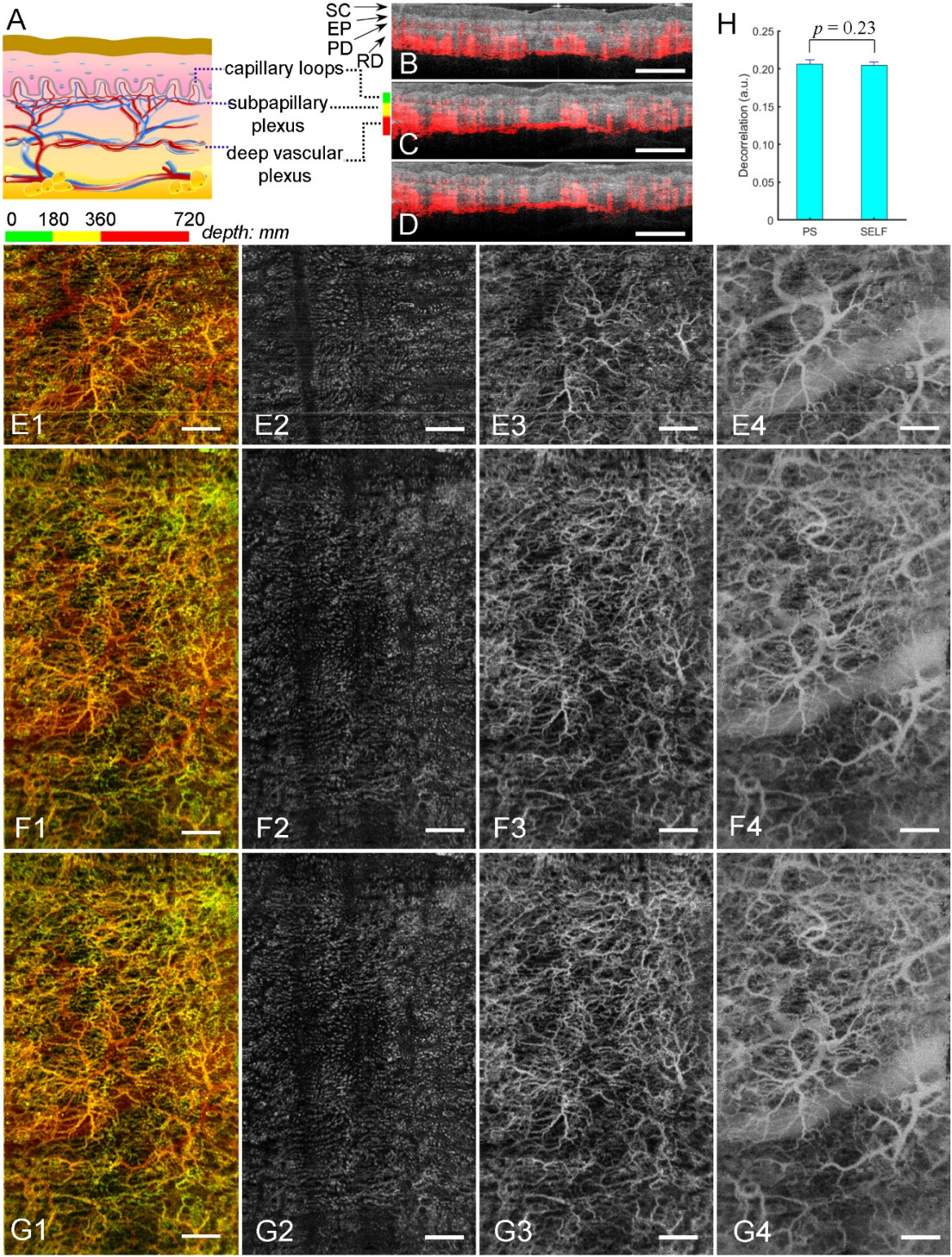

**Figure 3**. Comparison of FOV between the point scanning and SELF-OCTA with the same A-line rate and total acquisition time. (**A**) Schematic of the skin structure and vasculature. (**B**) A representative cross-sectional OCT structural image (gray) acquired using the point-scanning scheme overlaid with blood flow signals (red). SC: stratum corneum; EP, epidermis; PD, papillary dermis; RD, reticular dermis. (**C**&**D**) the corresponding images (gray) acquired using spectrally extended line field scheme before (C) and after (D) Y-deconvolution. The color bars in green, yellow and red correspond to the depth range of three *en face* slabs. (**E**) *En face* OCTA images acquired with the point-scanning scheme in the proximal interphalangeal joint of the middle finger (palm side) *in vivo*: **E1**, *En face* projection of three slabs color coded by depth, **E2**, the first slab mainly containing capillary loops, **E3**, the second slab mainly containing subpapillary plexus, **E4**, the third slab containing deep vascular plexus. (**F1-4**) *En face* SELF-OCTA images acquired at the same skin area corresponding to E1-E4 before Y-deconvolution. (**G1-4**) *En face* SELF-OCTA images corresponding to F1-4 after Y-deconvolution. (**H**). No statistically significant difference in decorrelation between the point-scanning (PS) and SELF-OCTA images. Scale bars: 1 mm.

There are always visible motion artifacts, appeared as thin bright and dark lines, in *en face* angiograms acquired with the point-scanning scheme (Figs. 3E1-4, Fig. 4A1-3, and **Supplementary Figs. 7A**). The corresponding SELF-OCTA images are almost free of such artifacts, because motion induced signal deviations are distributed into 16 Y image positions, substantially damping the contrast of motion artifacts (Figs. 3F1-4&G1-4, Fig. 4B1-3). This artifact damping mechanism is analogous to a selective low-pass filter along Y direction, which does not affect the signal. In a separate experiment, we deliberately generated motion artifacts by removing the vertical hand rest before image acquisition. Corresponding motion artifacts in SELF-OCTA *en face* images appear as low-intensity variations in the background (**Supplementary Fig. 6, Fig. 7B&C** and **Supplementary Movie 1**).

SELF-OCTA allows tailoring exposure time and inter-scan time without affecting FOV or total acquisition time. To validate this, we firstly acquired a 3D dataset with the point-scanning scheme with a nominal FOV of 6.55 mm x 6.55 mm and a total acquisition time of 3.28 s. The inter-scan time was 6.4 ms with an A-line rate of 80,384 Hz and 512 A-line per B-frame (Figs. 4A1-4). In the SELF scheme, we set the inter-scan distance to be 4Δx and the A-line rate to be 22,000 Hz, so that we were able to achieve 3.65 times longer inter-scan time and the same nominal FOV within 2.98 s (Figs. 4B1-3). Obviously, the advantage of longer inter-scan time is the significantly increased sensitivity to flow in small vessels and capillaries, which are largely invisible in the point-scanning OCTA images due to relatively shorter inter-scan time (Figs. 4A1-3, 4B1-3, 4D&4E). Notably, a practical advantage of longer integration time is ~10% larger X-scan duty cycle (Figs. 4A3&B3).

In addition, most SD-OCT devices are relative intensity noise (RIN) and electrical noise limited at working A-line rate. In the current device, total SNR is measured to be 9.94 dB lower at 80,384 Hz than that at 22,000 Hz A-line rate (Fig. 4C), which can be approximately broken down to 5.84-dB drop in signal due to reduced exposure time and 4.1-dB drop in signal to RIN ratio ($SNR_{RIN}$). This $SNR_{RIN}$ drop significantly elevates the noise background and overwhelms weak vessel signals from small vessels (Fig. 4A1-3&4D) compared with SELF-OCTA images (Fig. 4B1-3&4E).

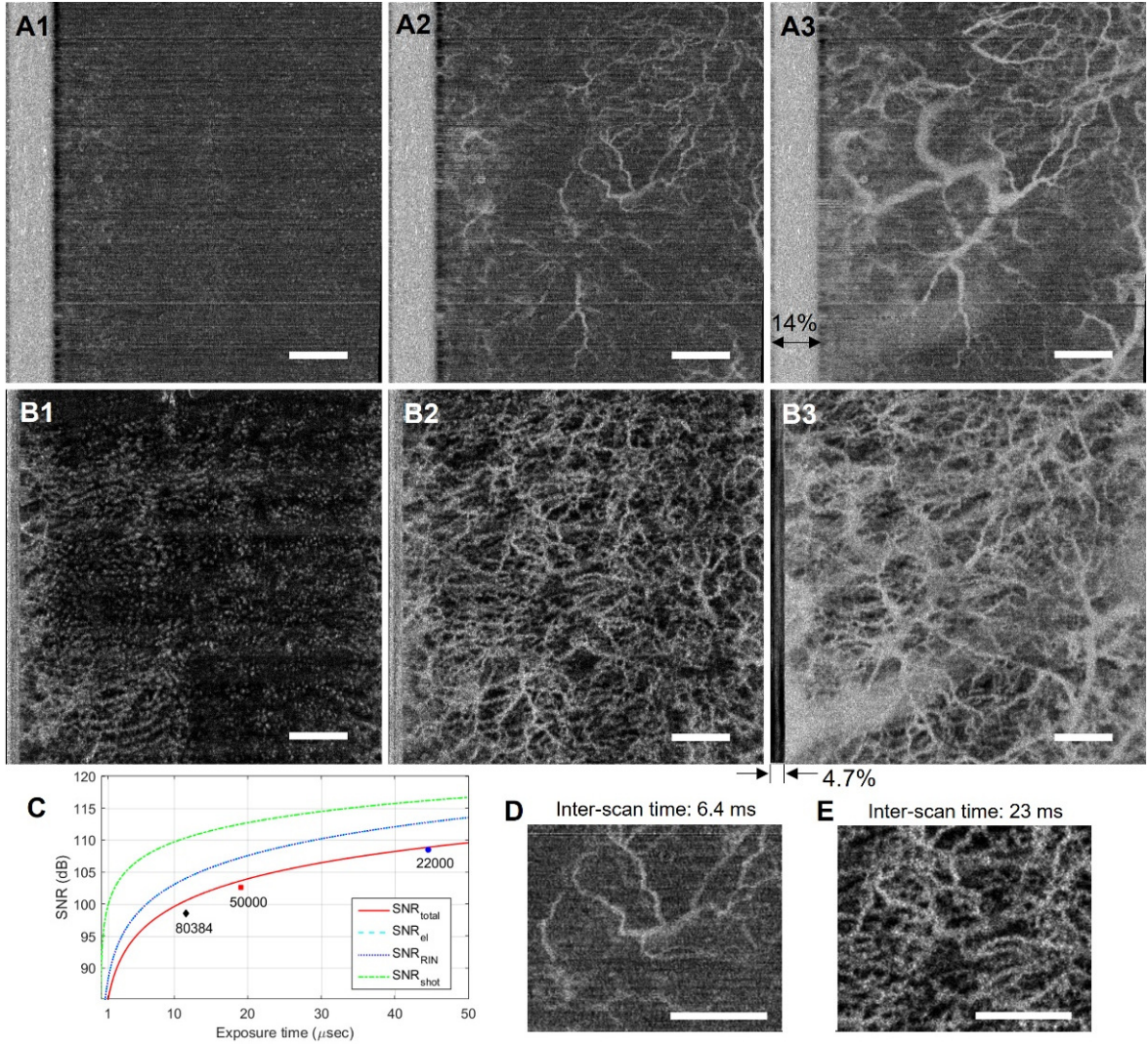

**Figure 4**. Comparison of vessel visibility and noise level between the point scanning and SELF-OCTA with the same input power and total acquisition time. (**A1-3**) *En face* OCTA projection of skin slab at the depth of capillary loop (A1), subpapillary plexus (A2) and deep vascular plexus (A3) acquired with the point-scanning scheme running at 80,384 Hz A-line rate. (**B1-3**) The corresponding SELF-OCTA images with *L* = 4 acquired at 22,000 Hz A-line rate. (**C**) Signal to noise ratio (SNR) as a function of exposure time. $SNR_{el}$, $SNR_{RIN}$, and $SNR_{shot}$, ratio of signal to electrical noise, relative intensity noise, and shot noise, respectively. Note that $SNR_{el}$ = $SNR_{rin}$ for optimal total SNR. Blue dot, red square and black diamond indicate measured SNR values, respectively. (**D&E**) Magnified views of a region of interest in A2 and B2. Scale bars: 1 mm.

Towards flow velocity quantification with high dynamic range, we have developed a partial B-frame scanning protocol that realizes multiple inter-scan time without increasing the number of B-scan repeats based on the SELF-OCTA platform. We split each B-scan of 384 A-lines into a half B-scan of 192 odd points (odd scan) and a half B-scan of 192 even points (even scan), so that there are 4 half-B-scans in a Y-scan cycle when *N* = 2 (Fig. 5A). OCTA images with inter-scan time of ΔT = 7.68 ms and ΔT/2 = 3.84 ms respectively can be achieved by re-arranging the order of half B-scans (Fig. 5A). Owing to the 'frequency flow' imaging mechanism, this protocol does not introduce any artifact since at each Y imaging position there are equal number of partial-spectrum decorrelation frames with inter-scan time of ΔT and ΔT/2. It is worth mentioning that this partial B-frame scanning protocol is not supported by the point-scanning platform since the inter-scan time, and consequently OCTA signal at odd and even pixels will be different.

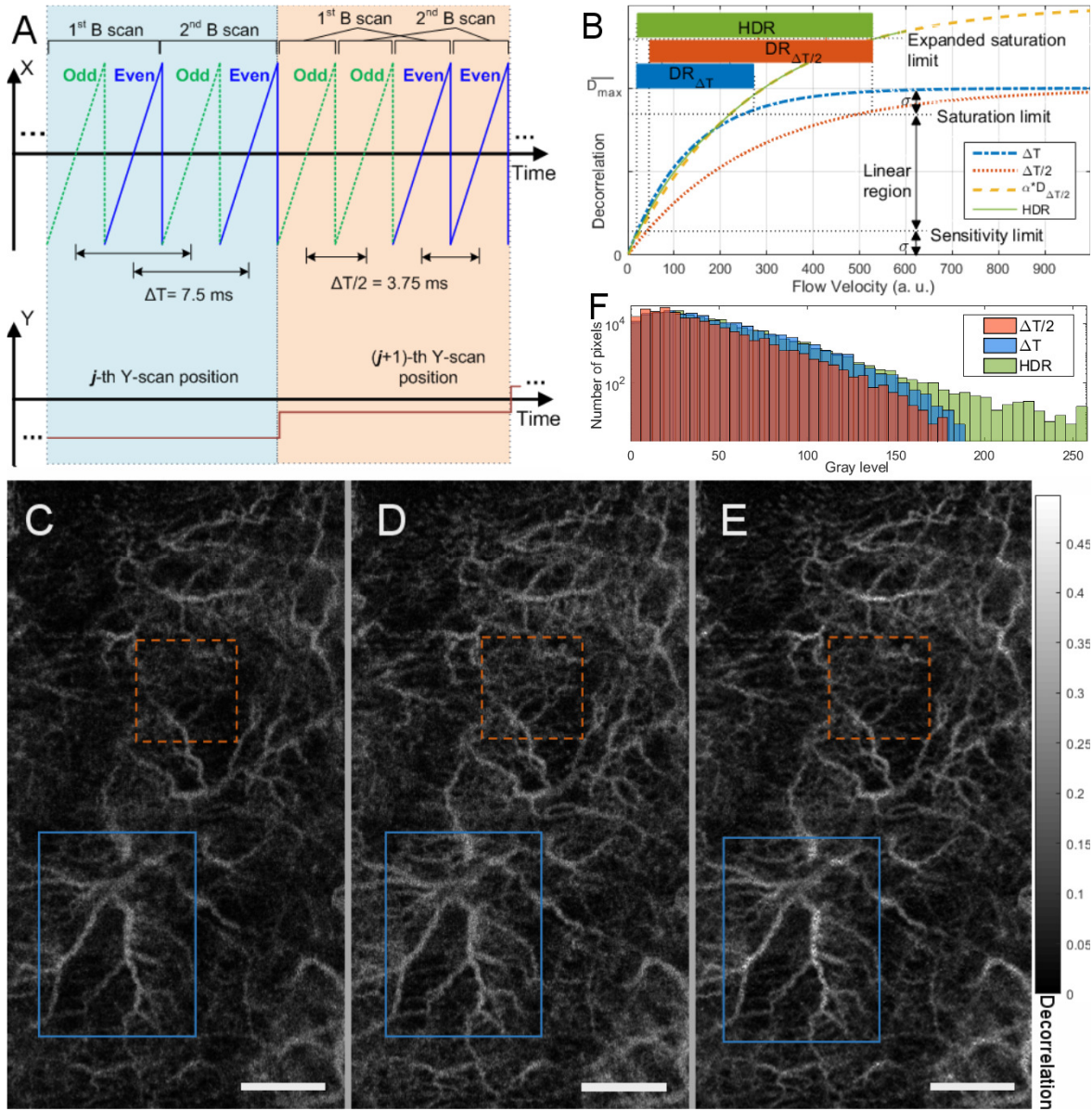

**Figure 5.** Multiple inter-scan time and high dynamic range (HDR) OCTA. (**A**) Waveforms of X-Y galvanometer scanner of two consecutive Y-scan cycles. Green dashed lines stand for odd points in a B-scan, and blue solid lines represent even points in a B-scan. (**B**) Model of decorrelation as a function of flow velocity with different inter-scan time. $\overline{D_{max}}$ is the mean saturated decorrelation signals, σ is the standard deviation of saturated decorrelation signals. $DR_{\Delta T}$ and $DR_{\Delta T/2}$ are the dynamic range of images acquired with inter-scan time of ΔT and ΔT/2. $D_{\Delta T}$ and $D_{\Delta T/2}$ refer to the decorrelation profiles of inter-scan time of ΔT and ΔT/2, respectively. α is the ratio between decorrelation signals acquired with ΔT ($D_{\Delta T}$) and ΔT/2 ($D_{\Delta T/2}$). (**C-E**) Representative *en face* angiograms acquired with ΔT/2 (C) and ΔT (D), and corresponding high dynamic range reconstruction obtained according to the model in B (E). Regions marked by orange dash-line boxes show vessels appearing silent with the ΔT/2 inter-scan time (C) but becoming visible with the ΔT inter-scan time (D). Regions marked by blue boxes show vessels having decorrelation signal differentiable with the ΔT/2 inter-scan time (C) but undifferentiable with the ΔT inter-scan time (D). (**F**) Histograms of image gray level acquired from C, D, and E. Scale bars: 1 mm.

For demonstration purpose, we introduce a simplified model to reconstruct angiograms with high dynamic range (Fig. 5B). Briefly, we assume that, over the decorrelation signal range of [σ, $D_{max}$-σ], flow velocities and decorrelation signals are linearly related, where $D_{max}$ is the mean saturation decorrelation and σ is the standard deviation of the saturation decorrelation. Based on this assumption, the ratio between decorrelation signals acquired with ΔT ($D_{ΔT}$) and ΔT/2 ($D_{ΔT/2}$) is a constant, α. If we multiply decorrelation profile of ΔT/2 with α, the new decorrelation profile, $α*D_{ΔT/2}$, has a higher saturation limit (Fig. 5B). The high dynamic range *en face* angiogram is reconstructed using both the angiogram acquired with ΔT and the angiogram corresponding to $α*D_{ΔT/2}$. Detailed process to generate the high dynamic range angiogram is provided in the ***Supplementary Information.*** Since this simplified model is based on a number of assumptions and approximations, readers are referred to previous phantom studies for an accurate model [16-18]. Nevertheless, this high dynamic range angiogram combines the merits of both inter-scan time (Figs. 5C-E): signals from small vessels with slow flow speed are retained which are otherwise invisible or silent in angiograms acquired with ΔT/2 (orange dash-line boxes, Fig. 5 C&E); signals from vessels with high flow velocities that are saturated in angiogram acquired with ΔT become approximately linearly related with the corresponding flow velocities (blue solid-line boxes, Fig. 5 D&E). The high dynamic range is also evident in the corresponding histograms (Fig. 5F).

To demonstrate motion tracking and correction, the subject deliberately moved the hand in the lateral directions during image acquisition. Since the piano wire attached with the skin blocks the light (Fig. 6A), in the *en face* angiograms the shadow appears bright due to high decorrelation of background noise (Figs. 6B, 6F&6G). There is a 14-pixel overlap in Y-axis between *en face* projections of 3D angiograms acquired in each two adjacent Y-scan cycles, as can be appreciated with the images of an air bubble in the refractive index matching gel (arrow, Fig. 6B). In a representative experiment, 2D lateral motions can be measured at the subpixel level throughout the whole FOV (Figs. 6C&D). The range of detected motion velocity was from 0-2.2 μm/ms (Fig. 6E). The performance of motion correction can be evaluated by comparing *en face* angiograms before (Fig. 6F) and after motion correction (Fig. 6G). The distorted piano wire image is largely corrected, which validates motion tracking and correction in the X direction. With an oblique direction of the piano wire with respect to Y-axis, the performance in both directions is demonstrated in ***Supplementary figs. 8&9***.

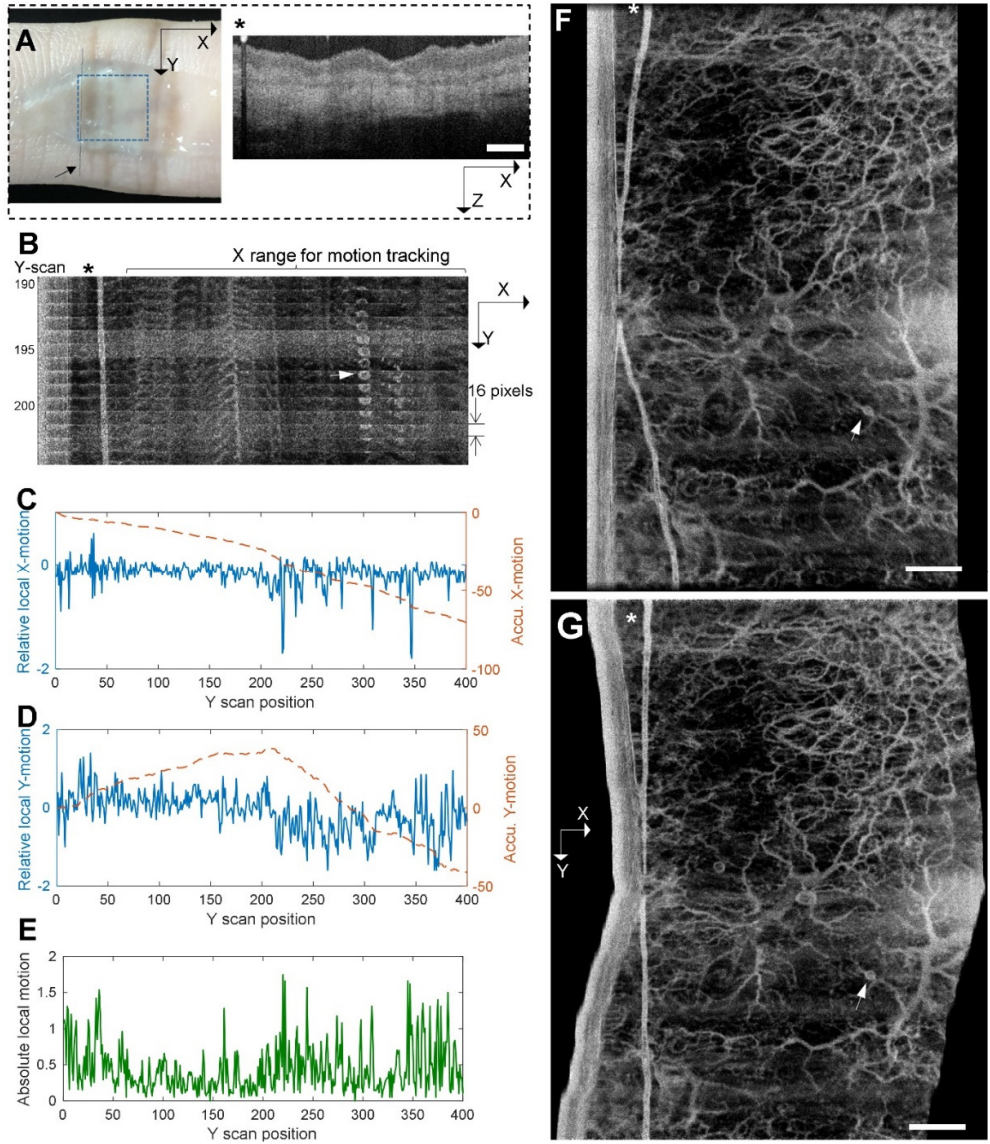

**Figure 6**. Motion tracking and correction in human skin *in vivo*. (**A**) (left) a photograph of the proximal interphalangeal joint of the middle finger (palm side) of a healthy human subject with a segment of piano wire (arrow) attached to skin surface covered by ultrasound transmission gel. Blue dashed box indicates imaging FOV. (right) A cross-sectional OCT structural image showing the shadow of the piano wire (star). (**B**) *En face* projections of 15 consecutive 3D angiograms with each acquired during a Y-scan cycle. The star indicates the position of piano wire. Images of an air bubble (arrow) in the gel demonstrate that the projects are overlapped. (**C**&**D**) Relative local (solid curve) and accumulative (accu.) motion (dashed curve) in X and Y-axis, respectively. The vertical unit is pixels. (**E**) Absolution local motion calculated as square root of sum of relative local motions squared. (**F**&**G**) A representative *en face* angiogram before (F) and after (G) motion correction, respectively. Arrows indicate the same air bubble as in B. Stars indicate the position of piano wire. Scale bars: 1 mm.

### Discussions

Unlike OCT structural imaging where cross-sectional viewing is preferred, OCTA is inherently three dimensional and typically displayed with *en face* projections. Therefore, 2D and 3D priority scanning mechanisms are superior to the conventional depth priority scanning mechanism in terms of imaging speed [28, 37-39]. As a 2D priority scanning platform, SELF-OCTA enables simultaneous signal acquisition from parallel lateral locations. This parallelization makes volumetric signal acquisition faster compared with the point-scanning scheme running with the same A-line rate. In achieving larger

FOV, an important feature of SELF-OCTA lies in that it has more sensitivity budget than the point-scanning scheme as higher MPE is allowed. In this study, we show a two-fold gain in FOV with no penalty in signal compared with the point-scanning scheme. Larger FOV can be readily achieved by increasing inter-scan distance independent of specifications of light source or signal detection hardware. Currently, ultrafast OCT systems are not clinically available, at least not to large patient populations. SELF-OCTA makes it possible for clinical OCTA devices to break their speed limit and achieve more than two times wider FOV by simply adding one optical element in the sample arm. Future translation of SELF-OCTA to ophthalmic imaging will fill the gap between OCTA and the standard-of-care tools in FOV, providing invaluable information about vascular changes that involve peripheral retinal regions.

Lack of flow velocity information in the current clinical systems is also associated with the imaging speed limitation as revealed by previous phantom studies [16-19]. Because of the above-mentioned trade-off between flow-signal linearity and sensitivity to slow flow, a wide range of inter-scan time from tens of microseconds to a few milliseconds are needed in order to cover the full range of retinal capillary flow from 0.4-3.0 mm/s [16, 19]. Current ultrahigh speed research prototypes provide an inter-scan time range of 1.5-4.5 ms, which are only applicable to flow velocity of <0.6 mm/s [16, 19]. Even this modest dynamic range requires at least three B-scan repeats [19]. Eventually, more B-scan repeats will be needed to extend the range to the higher flow speed, which will be at substantial cost of imaging speed, and consequently FOV. The 'frequency flow' mechanism offered by the SELF-OCTA realizes multiple inter-scan time independent of B-scan repeats. This unique capability overcomes the imaging speed bottleneck for velocimetry over large FOV. More importantly, since the inter-scan time of each partial-spectrum decorrelation frames can be adjusted independently over a wide range through the partial B-frame approach, the optimal trade-off between dynamic range and sensitivity of flow detection can be achieved. Therefore, SELF-OCTA will potentially overcome a wide variety of challenges where differentiation of flow velocity and degrees of flow impairment is of great clinically significance, for example, lack of flow velocity in flow index of large vessels for assessing optic disc perfusion [14] and difficulties in quantifying slow flow in microaneurysms [40].

The total spectral bandwidth determines the angular subtense, which further determines the MPE and the length of line field. The most restrictive scenario for SELF-OCTA will be posterior segment imaging centred at 1060 nm, where the maximum usable bandwidth is no more than 100 nm due to water absorption. The performance of SELF-OCTA will not be significantly affected by this limitation. For example, one can set the angular subtense to be 6 mrad, so that the corresponding length of line field at retina is 6 mrad x 17 mm = 102 μm. Setting the spacing between adjacent spectral band to be 7.7 μm and Hamming window length to be 40 μm (~40 nm), the spectrum can be split into 12 bands. Under these conditions, the axial resolution is estimated to be 16.5 um in tissue (refractive index = 1.38), and the lateral spot size in Y-axis is estimated to be 1.36 times larger than that of the monochromatic beam.

Deconvolution works well in restoring lateral resolution in Y-axis. Because deconvolution is an intensity-based model, there have been concerns when phase information is involved in the 3D coherent image formation [41]. Other concerns of using deconvolution in imaging scattering tissues are that the optical transfer function may not be accurate and that it is sensitive to speckle and noise [41]. In SELF-OCTA, artifacts associated with deconvolution are unnoticeable due to the following facts. First, the convolution involves phase information only in the depth dimension. Second, the speckle contrast is substantially reduced by frequency-temporal compounding as speckles of partial-spectrum decorrelation frames are fully uncorrelated (***Supplementary Fig. 10***). Thirdly, the optical transfer function is accurate since convolution is done digitally.

In the 'frequency flow' imaging mechanism, line-fields of adjacent Y-scan positions are overlapped in Y-axis, which for the first time makes it possible to track lateral motion with OCTA data. Current self-navigation motion correction method can only provide an indicator that motion occurred, and lack of quantitative lateral motion trajectory information results in artifacts such as vessel interruptions and low sensitivity of ocular drift [11]. SELF-OCTA may complement the self-navigation method. The typical angular velocity of ocular drift during fixation is 1 °/s, which corresponds to a translational retinal motion of 0.17 µm/ms if the radius of retina is 10 mm. Such slow motions should be readily detected with SELF-OCTA as shown in Fig. 6. The overlap between scanned areas of adjacent Y-scan cycles may be used to correct vessel interruptions. Combination of SELF-OCTA and self-navigation motion correction method will potentially obviate the problems with hardware-based tracking systems.

Currently, SD-OCT is the most widely used system, mainly due to the matured technology, excellent phase stability for functional imaging, and lower cost. In SD-OCT, RIN is inversely proportional to exposure time, prohibiting this well-established technology from high-speed applications, particular when compared with SS-OCT where RIN is largely suppressed by dual-balanced detection [6]. As a result, the speed of SD-OCT clinical devices is limited to 70k Hz in contrast to 100-200 kHz of SS-OCTA [2]. SELF-OCTA disassociates imaging speed with exposure time, which overcomes this standing problem and will make SD-OCTA viable for high-speed OCTA applications.

In conclusion, SELF-OCTA platform facilitates multiple important technological advancements towards wide field, quantitative, and low-cost angiographic imaging. SELF-OCTA breaks the speed limitation without sacrificing sensitivity, so that wide field imaging will be realistic in devices with speed limitation, in particular, SD-OCT. SELF-OCT also overcomes the speed bottleneck of quantitative flow velocity imaging, making this important function available with little cost. Last but not the least, OCTA-data based motion tracking capability may potentially simplify the complexity and reduce the cost of OCTA devices. In principle, by simple modifications in the sample arm, all the existing clinical devices, regardless of type and A-line rate, should acquire these advanced imaging capabilities without affecting existing functions. We expect that SELF-OCTA will make wide field, quantitative OCTA an inexpensive technology, therefore, widely available to larger patient populations.

**Acknowledgement**

We acknowledge the funding support from the Singapore Ministry of Health's National Medical Research Council under its Open Fund Individual Research Grant (MOH-OFIRG19may-0009), and the Ministry of Education Singapore under its Academic Research Funding Tier 2 (MOE-T2EP30120-0001).

**Author contributions**

L.L. conceived the project. S.C. developed the imaging system. K.L. and S.C. participated in software development. S.C. and L.L. performed image acquisition and data analysis. L.L. and S.C. wrote the manuscript. All authors edited the final version of the paper.

**Spectrally Extended Line Field Optical Coherence Tomography Angiography**

Si Chen, Kan Lin, and Linbo Liu*

*School of Electrical & Electronic Engineering, Nanyang Technological University, Singapore*

*Correspondence: L. L. (liulinbo@ntu.edu.sg)

**Supplementary Information**

**1. Imaging system construction and characterization**

We combine two superluminescent diode modules (IPSDS1313 and IPSDS1201C, Inphenix, CA, USA) with a 50:50 fibre coupler (TW1300R5A2, Thorlabs Inc, USA) (Fig. 1). The combined light source provides a radiation from 1230 nm to 1360 nm (-6 dB). The outputs of the fibre coupler are connected to two optical circulators (PIBCIR-1214-12-L-10-FA, FOPTO, Shenzhen, China), which guide the light beams to the sample arm and reference arm, respectively. The light back reflected from the reference arm and back-scattered from the sample arm are combined using a 95:5 fibre coupler (WP3105202A120511, AC Photonics, CA, USA). In the sample arm of the point scanning scheme, the sample beam is firstly collimated by an achromatic lens L1 (AC050-010-C, Thorlabs Inc., USA) before reflected by a mirror (RM) and a pair of galvanometer scanners and focused by an objective lens L2 (AC254-050-C, Thorlabs Inc, USA).

In the SELF-OCTA scheme, the mirror (RM) is replaced by a set of three identical prisms (N-SF11, PS872-C, Thorlabs Inc.) with apex angle of 30° and angular spacing of ~56.9°. Switching between the point scanning scheme and SELF-OCTA scheme is realized with a manual translation stage (Fig. 1A). The prism is with anti-reflection coating so that one-way transmission efficiency of three prisms was measured to be 94%. The polychromatic sample beam is dispersed by the prisms into a line along the slow (Y) axis in the focal plane of the objective lens L2 (Fig. 1B). The spectrometer is comprised of a collimating lens L5 (AC254-035-C, Thorlabs Inc., USA), a transmission grating (PING-sample-106, Ibsen Photonics, Denmark), a home-made multi-element camera lens and a line scan camera (LDH2, Sensors Unlimited, USA). The camera pixel size is 25 µm by 500 µm (width by height) and we used all 1024 pixels. The total spectrometer efficiency was measured to be 0.61, including the quantum efficiency of the camera. The spectral resolution is 0.148 nm, resulting in a total ranging depth of 2.89 mm in air. The axial resolution was measured to be 9.82 µm in air. The 6-dB ranging depth was measured to be 1.6 mm in air and the sensitivity roll-off over depth is ~3.75 dB/mm. With the optical power incident on the sample being 4.74 mW, the sensitivity measured at ~150 µm from DC is 108.52 dB, 102.56 dB, and 98.58 dB at the A-line rate of 22k Hz, 50k Hz, and 80k Hz, respectively, which agree with the theoretical predictions (Fig. 4C).

We used 16 Hamming windows with size of 263 pixels and spacing of 52 pixels to generate 16 spectral bands (Supplementary Fig. 2A), respectively. The axial resolution (FWHM) of each band was measured to be ~28.5 µm in tissue (refractive index = 1.38) (Fig. 2C). The transverse resolution (FWHM) of each band was measured to be 39.6 µm (10-90% width of an edge scan) because of the convolution between the Hamming window and monochromatic point-spread function (PSF) (Supplementary Fig. 2B), which is 1.51 times larger than the monochromatic transverse resolution and agree well with theoretical prediction (Supplementary Fig. 2C). The trade-off between axial and transverse resolution along Y-axis is characterized in Supplementary Fig. 2D.

The theoretical transverse spot size at 1310 nm is ~27 µm (full-width at half maximum, FWHM) since the nominal mode field diameter of SMF-28e fibre is 9.2 µm at 1310 nm. The monochromatic transverse resolution at 1310 nm was approximated to be 26.2 µm by measuring 10-90% width of an edge scan using the part of signal at the centre of the spectrum with a narrow line width (~1.5 nm

FWHM) (Fig. 2D). For a Gaussian spot, the lateral resolution, defined at its $e^{-2}$ radius, can be shown to be 0.78 times the 10–90% edge width [42], so that the monochromatic spot size (FWHM) is estimated to be 24.1 µm.

For deconvolution along Y direction, we used Lucy-Richardson method (deconvlucy in MATLAB®). We used the Hamming window mentioned above as the point-spread function (PSF in deconvlucy) and damping threshold of 2.

## 2. Maximum permissible exposure of line field

Following the literature [35], the relevant 'most restrictive' evaluation angular subtense is the one with the maximum ratio of partial power/ evaluation angular subtense. The partial power is the total optical power contained in the evaluation angular subtense. The intensity profile of the line field was measured by scanning a fiber tip (SMF-28e) along Y-axis in the focal plane of the objective lens and record the optical power coupled into the fiber (Supplementary fig. 1A). Since it is a line illumination, the angular subtense along X-axis takes 1.5 mrad [34]. Therefore, the ratio of partial power/ evaluation angular subtense

$$\frac{P}{\delta} = 2P/(\alpha_Y + 1.5\ mrad) \qquad (1)$$

According to the equation (1), the angular subtense in Y-axis ($\alpha_Y$) obtained using 'Most restrictive ratio' analysis is 3.176 mrad (Supplementary fig. 1B). The extended source correction factor $C_E$ is (3.176 +1.5 mrad) /2/1.5 mrad = 1.5587. Note that power limit calculated using $C_E$ applies only to the partial power within the angular subtense δ, instead of the total power. There are 20.11% of total power that is outside the angular subtense δ. For the experiments conducted with the point-scanning scheme and total image acquisition time of 4.096 s (Fig. 3 and Supplementary Fig. 8), the optical power incident on the sample is 4.74 mW. The corresponding power for the SELF scheme is calculated as (4.74 mW × $C_E$) / (1 - 0.2011) = 9.25 mW. Therefore, we used ~9.10 mW for all the experiments assuming the SELF scheme. For the experiments conducted with the point-scanning scheme and total image acquisition time of 3.26 s, the optical power incident on the sample is 9.1 mW (Figs. 4A1-3).

## 3. Image acquisition

The sample beam focused by the objective lens L2 propagates along the horizontal direction to the sample (Fig. 1B). We used a compact lab jack (L200, Thorlabs Inc.) as the vertical hand rest and a cage plate (CP33T/M, Thorlabs Inc.) as the horizontal hand rest to minimize motion. We applied ultrasound transmission gel (Aqua Sonic 100) to the skin area to be imaged to avoid high reflection signal from the skin surface.

Most of the images are acquired at 50k Hz A-line rate, except for Fig. 4 in which 22k Hz and 80 k Hz are used. There are 512 A-lines per B-scan, except for Fig. 5 where 384 A-lines per B-scan is used. For images generated with the point-scanning scheme, there are 400 cross-sectional angiograms so that the field of view (FOV) is 6.55 mm by 5.12 mm (width by height). For images generated with SELF-OCTA and **L** = 2, there are 800 cross-sectional angiograms so that the FOV is 6.55 mm by 10.24 mm (width by height). For all the OCTA images in this study, the number of repeated B-scans **N** is 2.

## 4. OCT structural image processing

Owning to the 'frequency flow' scanning mechanism, there are **M/L** partial-spectrum decorrelation frames at each Y image position, acquired during **M/L** consecutive Y-scan cycles. Note that **N** = 2 repeated B-scans at a Y-scan position are considered as one complete Y-scan cycle. We corrected bulk motion among all the partial-spectrum interference data before coherently added them into a spectrum interference data with the full spectral bandwidth. The phase differences between B-scans of consecutive Y-scan positions are calculated following the method described by An *et al* [20]. The axial resolution of SELF-OCT cross-sectional images are comparable to that of the point-scanning OCT (Supplementary Fig. 4). However, due to residual bulk motion and local tissue motion such as blood flow, SELF-OCT cross-sectional images are not as crispy as that of the point-scanning OCT.

## 5. Speckle contrast analysis

The speckle contrast is calculated using the following equation introduced in a literature [39]. We additionally analyzed speckle contrast of *en face* SELF-OCTA images with 1, 2, 4, and 8 partial-spectrum decorrelation frames at each Y image position, which corresponds to the cases of **L** = 16, 8, 4 and 2, respectively.

$$Speckle\ contrast = \frac{\sigma_S^2}{\overline{I_S}} \qquad (2)$$

Where $\overline{I_S}$ is the averaged intensity over the area of signal, and $\sigma_S$ is the standard deviation of signal intensity. We selected 24 regions of interest (ROIs) in a blood vessel with most of the pixels saturated (maximum decorrelation) as shown in Supplementary Figs. 10A-D.

## 6. Multiple inter-scan time and high dynamic range

The model of dynamic range extension is based on the autocorrelation equation of the laser speckle signal [19, 43]

$$\overline{D} = 1 - exp\left[-\frac{\tau \cdot (v_{flow} + v_{brownian} + v_{bulk})}{k}\right] \qquad (3)$$

Where $\overline{D}$ is the amplitude decorrelation, $\tau$ is the exposure time, which can be correlated to inter-scan time ΔT in OCTA. $v_{flow}$ is the velocity of the blood flow, $v_{brownian}$ and $v_{bulk}$ are velocity related with Brownian and bulk motion of the tissue. *k* is a constant.

According to this simplified equation (3), the decorrelation functions corresponding to the inter-scan time of ΔT and ΔT/2 are plotted in Fig. 5B. Obviously, there is a range of flow velocity over which the decorrelation is approximately linearly dependent on flow velocity. Following the method described in the literature [19], we define the decorrelation range of [σ, $\overline{D_{max}}$-σ] to be the linear region, where σ is the standard deviation of decorrelation signals. That is to say, σ is the sensitivity limit and $\overline{D_{max}}$-σ is the saturation limit. The dynamic ranges of signals acquired with ΔT and ΔT/2 are then defined as the flow velocity ranges corresponding to the linear region, respectively.

Since in the linear region, decorrelation signals acquired with ΔT and ΔT/2 are assumed to be linear, the ratio between their slopes α can be obtained. If we multiply the decorrelation profile of ΔT/2 with α, the new signal, referred as α*D_{ΔT/2}, overlap with decorrelation signal acquired with ΔT in the linear region. Since this simplified model is based on a number of assumptions and approximations, readers are referred to previous phantom studies for an accurate model [16-18].

According to the model proposed above, the procedures to construct high dynamic range angiograms are listed below:

1. Measure the mean maximum decorrelation signal ($\overline{D_{max}}$ =0.2079) and the standard deviation σ (σ =0.0151) of decorrelation signal from blood vessels in the amplitude decorrelation image of inter-scan time ΔT;

2. Perform 3x3 median filtering on the *en face* SELF-OCTA images acquired with ΔT and ΔT/2. We term the new images ΔT_median and ΔT/2_median, respectively;

3. Remove all pixels in the *en face* SELF-OCTA images acquired with ΔT, whose decorrelation value in ΔT_median is outside the linear region [σ, $\overline{D_{max}}$ -σ]. The new image is named ΔT_linear. Remove all pixels in the *en face* SELF-OCTA images acquired with ΔT/2, whose decorrelation value in ΔT/2_median is outside the linear region [σ, $\overline{D_{max}}$ -σ]. The new image is named ΔT/2_linear.

4. Find α using histogram of ΔT_linear – α*ΔT/2_linear. The histogram is centred at 0 when α = 1.35.

5. The high dynamic range image is synthesized as follows (green curve in Fig. 5B). For pixels in the linear region, the decorrelation values take the mean of ΔT_linear and α*ΔT/2_linear; for pixels with value larger than $\overline{D_{max}}$ –σ, the decorrelation will take the corresponding values in α*$D_{ΔT/2}$; for pixels with value smaller than σ, the decorrelation will take the corresponding values in ΔT.

**7. Motion tracking and correction**

Transverse motion is calculated by use of 2D registration algorithm available via:

http://mtshasta.phys.washington.edu/website/superSegger/SuperSegger/Internal/dftregistration.html

This registration algorithm returns relative motion in pixels along the two transverse directions between 2D *en face* projects of adjacent Y-scan positions, which is also termed as local motion. The accumulative motion is then used for correcting motion between partial spectrum decorrelation frames at each Y imaging position.

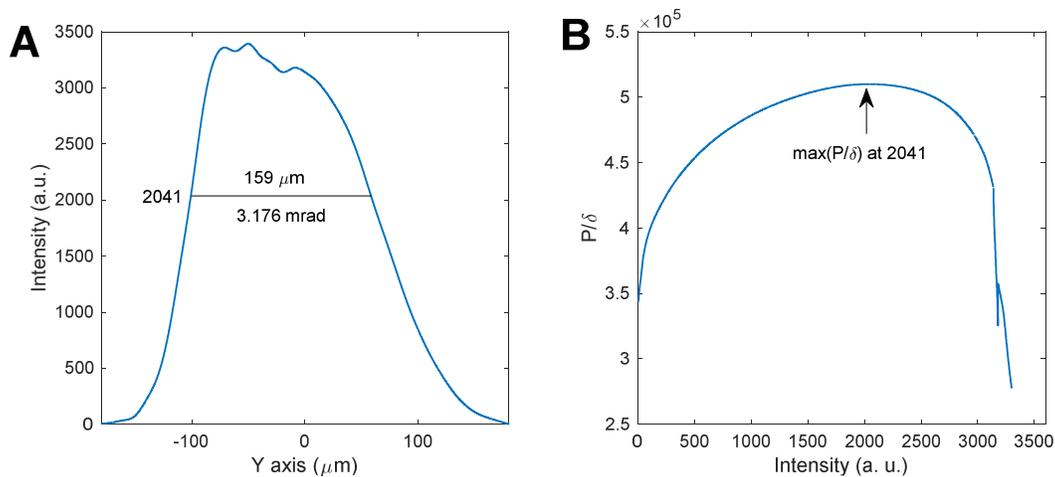

**Supplementary Fig. 1.** (**A**) Intensity profile of focus along Y-axis. (**B**) Ration between partial power (P) within window width (δ) for "Most Restrictive Ratio' method. The window width is 159 μm (3.176 mrad) when the ratio reaches the maximum.

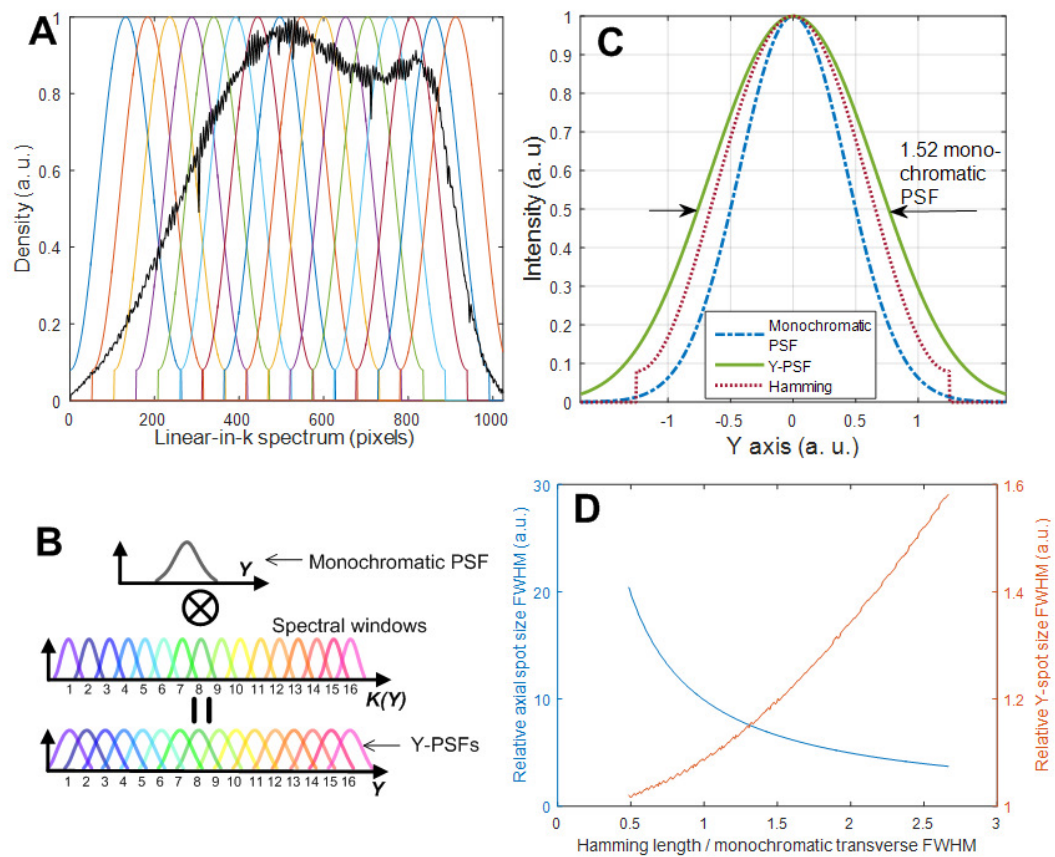

**Supplementary Fig. 2**. (**A**) Spectrum detected by the OCT spectrometer in *k*-space and Hamming filters. (**B**) Spectral convolution process. (**C**) Transverse point-spread function (PSF) along Y-axis (Y-PSF) as the result of convolution between the Hamming window and monochromatic PSF. (**D**) Spatial resolution as a function of hamming window length. Relative axial spot size = axial resolution of each spectral band / axial resolution of full spectrum. Relative Y-spot size = transverse resolution along Y-axis / monochromatic transverse resolution.

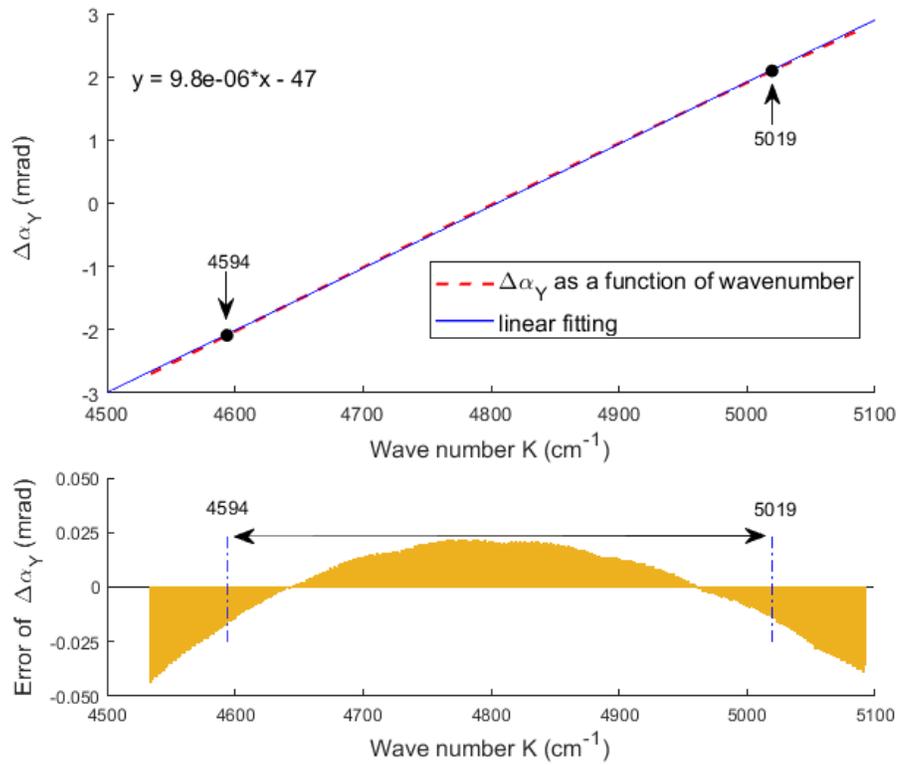

**Supplementary Fig. 3**. (**A**) Angular substance along Y-axis ($\Delta\alpha_Y$) as a function of wave number (K) (red dash line). Blue solid line is the linear fitting. The center wave number of the 1st and 16th spectral band are 4594 cm$^{-1}$ and 5019 cm$^{-1}$, respectively. (**B**) Linear error of $\Delta\alpha_Y$ is <0.0217 mrad within the range from 4594 cm$^{-1}$ to 5019 cm$^{-1}$, corresponding to <1.0855 µm in the focal plane of the objective lens.

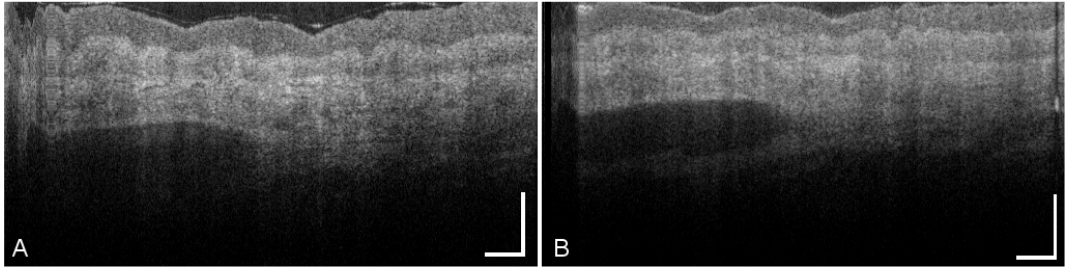

**Supplementary Fig. 4**. Cross-sectional structural images (amplitude frames) of human skin obtained using the point-scanning OCT (**A**) and SELF-OCT (**B**). Scale bars: 0.5 mm.

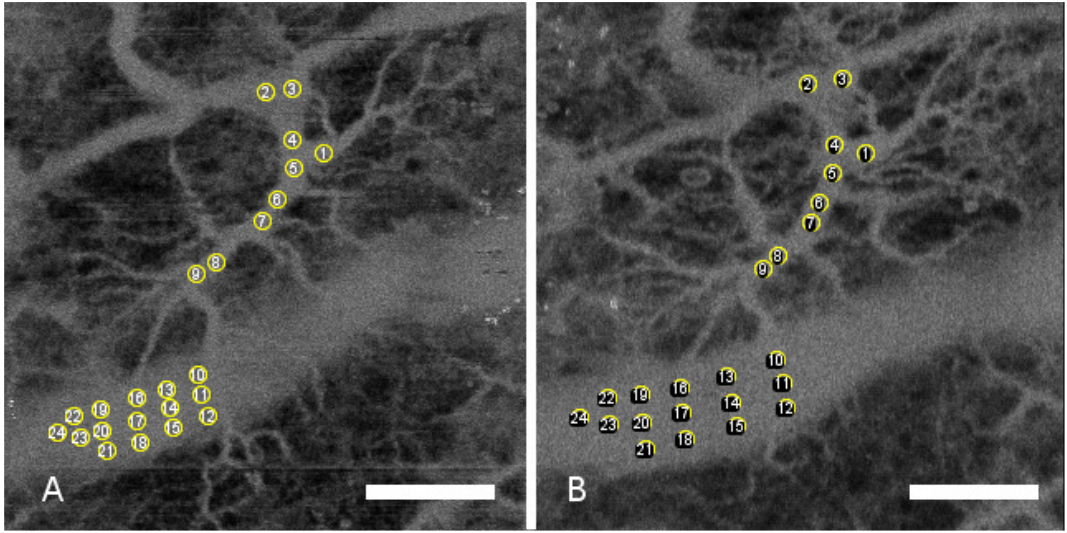

**Supplementary Fig. 5**. ROIs for comparison of decorrelation between point-scanning (**A**) and SELF OCTA (**B**). Scale bars: 1 mm.

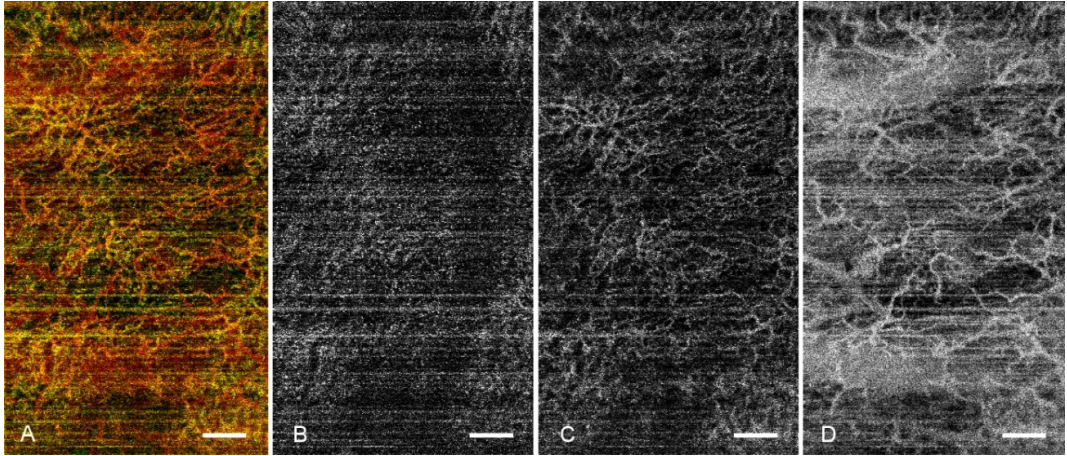

**Supplementary Fig. 6**. Representative *en face* SELF-OCTA projections reconstructed from the dataset with one partial-spectrum decorrelation images per Y-image position. (**A**) color coded projection of three slabs, (**B**) slab of capillary loop, (**C**) slab of subpapillary plexus and (**D**) slab of deep vascular plexus. Scale bars: 1 mm.

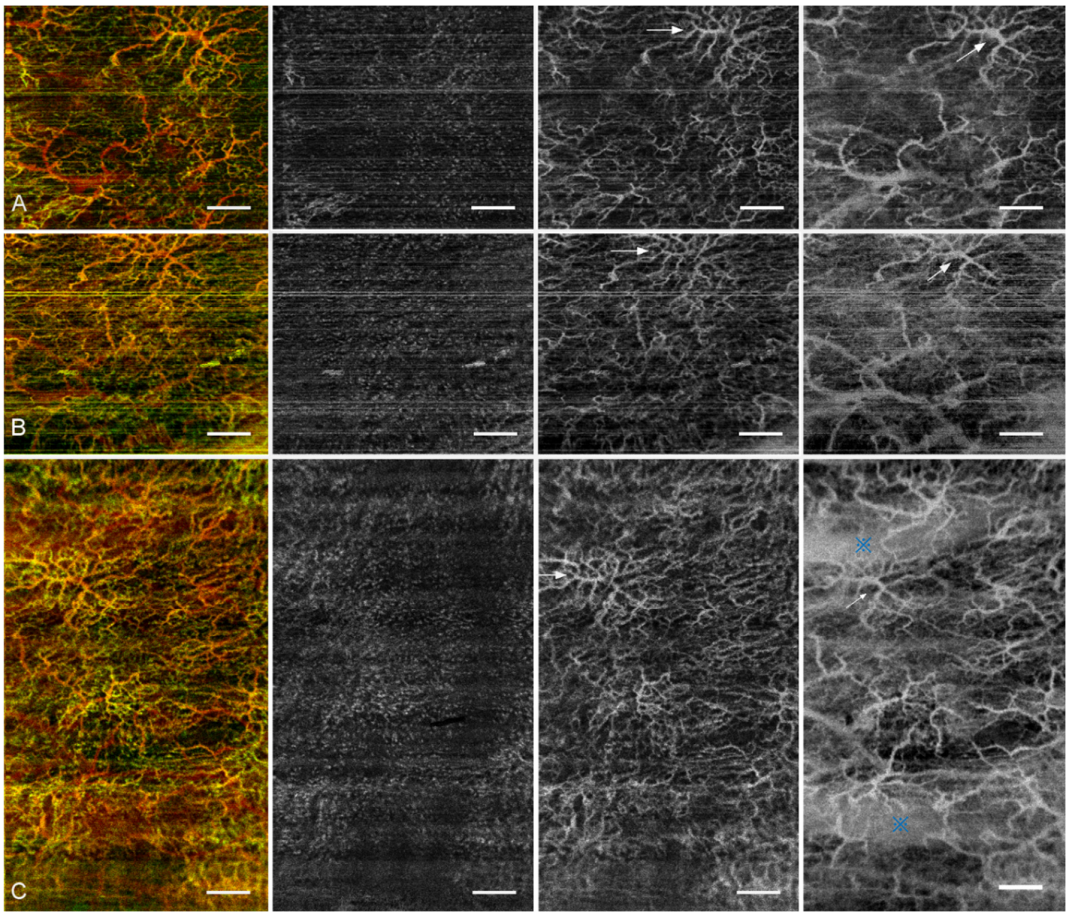

**Supplementary Fig. 7**. (**A**) *En face* angiograms generated with the point-scanning scheme. Motion was minimized using both horizontal and vertical hand rests during image acquisition. (**B**) *En face* angiograms produced by the point-scanning scheme. Motion was due to lack of the vertical hand rest during image acquisition. (**C**) *En face* angiograms generated with the SELF-OCTA scheme, Motion was due to lack of the vertical hand rest during image acquisition. Arrows indicate the same blood vessel. Blue stars indicate large blood vessels at the deep layer of the skin. Scale bars: 1 mm.

39

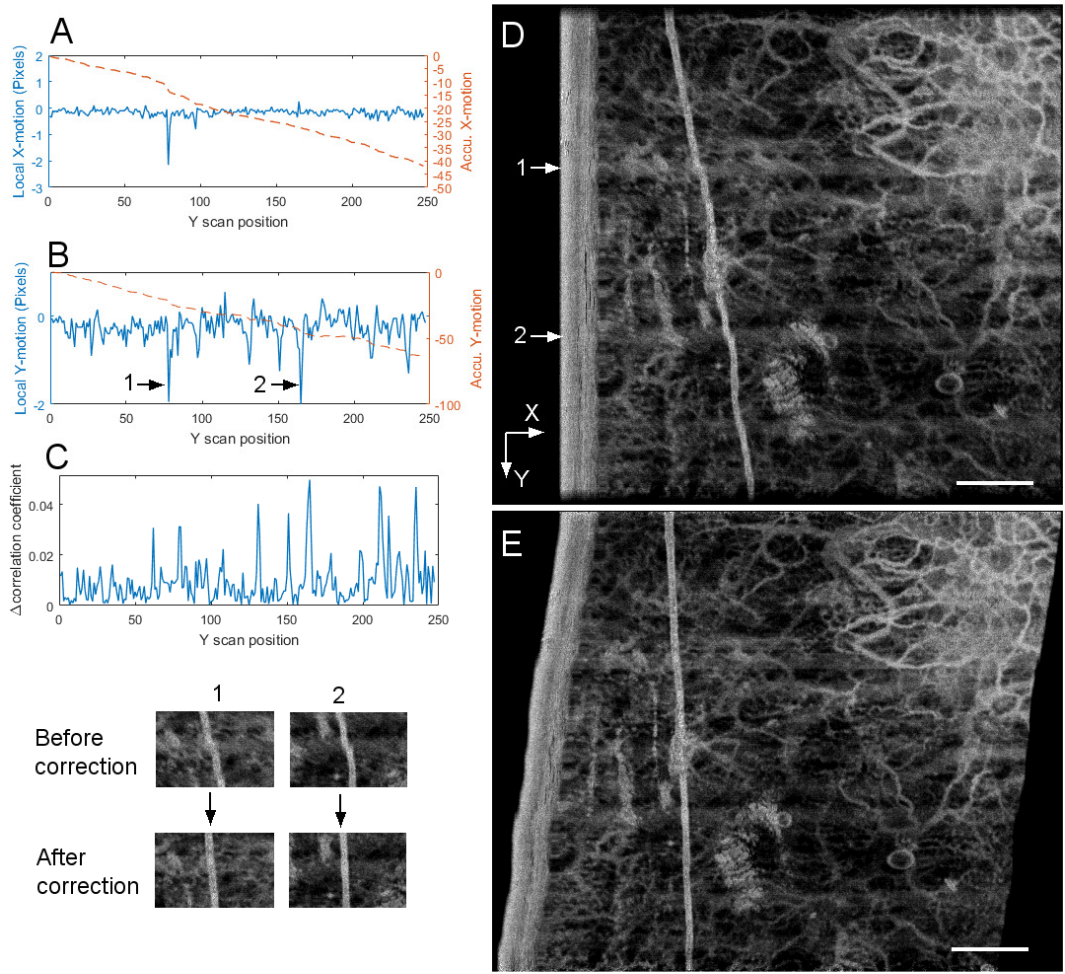

**Supplementary Fig. 8**. SELF-OCTA based motion tracking and correction. (**A**)&(**B**) Registered motion along X and Y-axis, respectively. Local motion stands for relative motion between 2D *en face* projections acquired in a complete Y-scan cycle. Accu. X/Y-motion refers to the total motion accumulated at a Y imaging position relative to the first B-scan. (**C**) Δ correlation coefficient is the difference in correlation coefficient between adjacent 2D *en face* projections before and after correction. (**D**)&(**E**) En face angiograms before and after motion correction, respectively. (Lower left) Regions (1,2) with local motion = 2 pixels before and after correction.

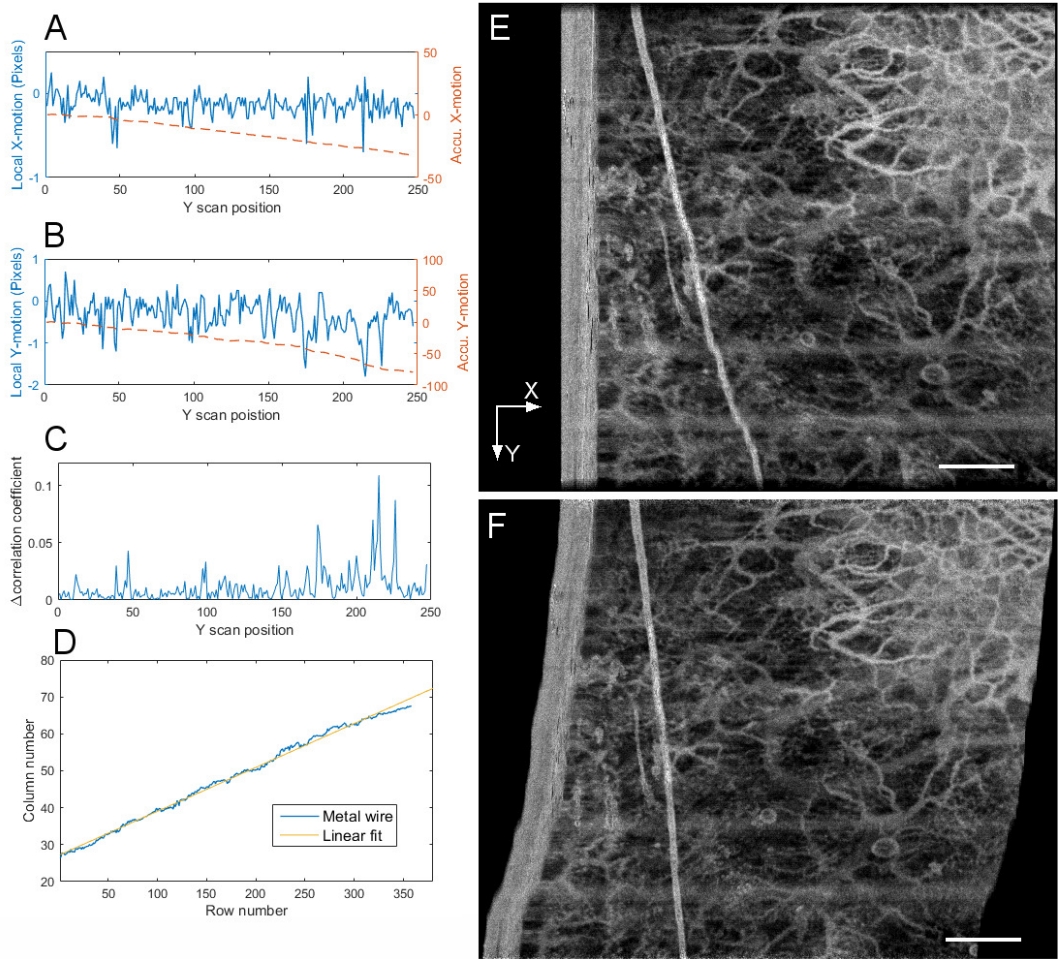

**Supplementary Fig. 9**. SELF-OCTA based motion tracking and correction. (**A**)&(**B**), Registered motion along X and Y-axis, respectively. Local motion stands for relative motion between 2D *en face* projections acquired in a complete Y scan cycle. Accu. X/Y-motion refers to the total motion accumulated at a Y imaging position relative to the first B-scan. (**C**) Δ correlation coefficient is the difference in correlation coefficient between adjacent 2D *en face* projections before and after correction. (**B**) Metal wire position and its linear fit. With its linear fit as the mean, the standard deviation of the wire transverse position is 0.8537 pixels, which is equivalent to 10.9 μm. (**E**)&(**F**) *En face* angiograms before and after motion correction, respectively.

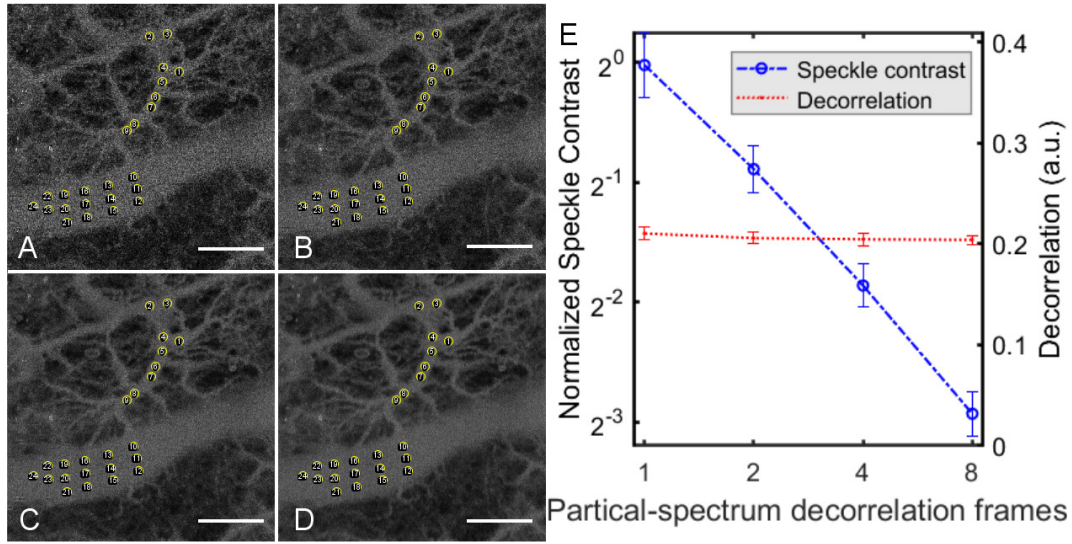

**Supplementary Fig. 10**. Speckle contrast in *en face* SELF-OCTA images dependent on the number of partial-spectrum decorrelation frames at each Y image position. (**A-D**) ROIs for speckle contrast with averaging 1 partial-spectrum decorrelation frame (3$^{rd}$ frame, **A**); averaging 2 partial-spectrum decorrelation frames (3$^{rd}$ and 7$^{th}$ frames, **B**); averaging 4 partial-spectrum decorrelation frames (1$^{st}$, 3$^{rd}$, 5$^{th}$, and 7$^{th}$ frames, **C**); averaging 8 partial-spectrum decorrelation frames (all 1$^{st}$ to 8$^{th}$ frames, **D**). (**E**) Speckle contrast of SELF-OCTA signals is inversely proportional to the number of partial-spectrum decorrelation frames in frequency-temporal compounding. Scale bar: 1mm.

**Supplementary Movie 1**. Sequential display of *en face* SELF-OCTA projections with each reconstructed from a 3D dataset of one partial-spectrum decorrelation frame. The number at the bottom indicate the sequence number of the partial-spectrum decorrelation frame.